\documentclass{vgtc}                          

\ifpdf
  \pdfoutput=1\relax                   
  \usepackage{graphicx}                
  \DeclareGraphicsExtensions{.pdf,.png,.jpg,.jpeg} 
\else
  \usepackage{graphicx}                
  \DeclareGraphicsExtensions{.eps}     
\fi%

\graphicspath{{figures/}{pictures/}{images/}{./}} 

\usepackage{microtype}                 
\PassOptionsToPackage{warn}{textcomp}  
\usepackage{textcomp}                  
\usepackage{mathptmx}                  
\usepackage{times}                     
\usepackage{cite}                      
\usepackage{tabu}                      
\usepackage{booktabs}                  

\onlineid{3203}

\vgtccategory{System}

\vgtcinsertpkg

\title{Reality Distortion Room: A Study of User Locomotion Responses to Spatial Augmented Reality Effects\thanks{This is a preprint version of this article. The final version of this paper can be found in the Proceedings of IEEE ISMAR 2023. For citation, please refer to the published version. This work was initially made available on the Microsoft Research website [microsoft.com] on September 2023, and was subsequently uploaded to arXiv for broader accessibility.}}

\author{You-Jin Kim\\ %
    \parbox{1.4in}{\scriptsize \centering University of California \\ Santa Barbara} %
\and Andrew D. Wilson\\ %
    \scriptsize Microsoft Research %
\and Jennifer Jacobs\\ %
    \parbox{1.4in}{\scriptsize \centering University of California \\ Santa Barbara} %
\and Tobias Höllerer\\ %
    \parbox{1.4in}{\scriptsize \centering University of California \\ Santa Barbara}}

\teaser{
  \centering
  \includegraphics[width=\linewidth]{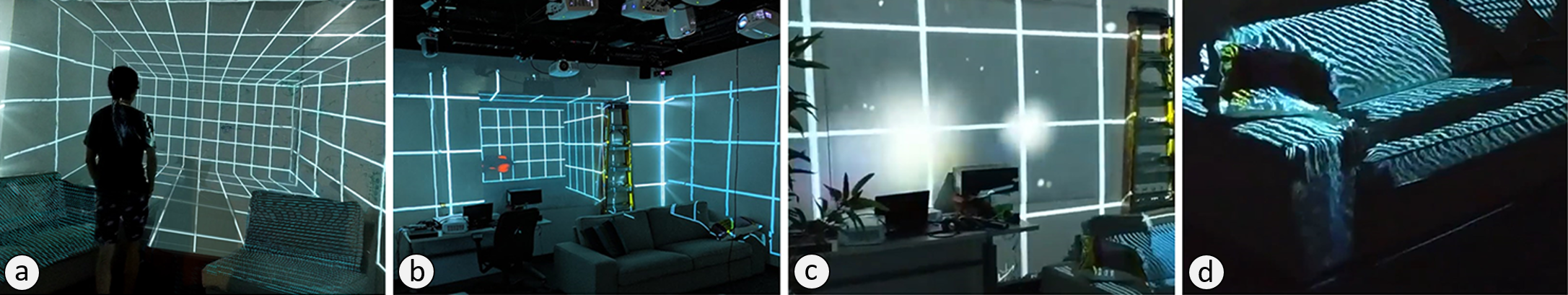}
  \caption{The Spatial Augmented Reality (SAR) environment is augmented by four depth cameras and five short throw overhead projectors. a. User looking at a wall as the room's geometric shape transforms. b. First-person point of view (POV) from the perspective of a user viewing a virtually extended space in the room. c. To improve users' spatial awareness, numerous floating particles were rendered to their POV in the environment. d. In order to ensure user safety in a relatively dark area, texture pattern overlays are projected onto physical furniture to make it look like an illuminated hologram-like item.}
  \label{fig:teaser}
}

\abstract{Reality Distortion Room (RDR) is a proof-of-concept augmented reality system using projection mapping and unencumbered interaction with the Microsoft RoomAlive system to study a user’s locomotive response to visual effects that seemingly transform the physical room the user is in. This study presents five effects that augment the appearance of a physical room to subtly encourage user motion. Our experiment demonstrates users’ reactions to the different distortion and augmentation effects in a standard living room, with the distortion effects projected as wall grids, furniture holograms, and small particles in the air. The augmented living room can give the impression of becoming elongated, wrapped, shifted, elevated, and enlarged. The study results support the implementation of AR experiences in limited physical spaces by providing an initial understanding of how users can be subtly encouraged to move throughout a room.%
} 

\CCScatlist{
  \CCScatTwelve{Human-centered computing}{Empirical studies in HCI}{}{};
  \CCScatTwelve{Computing methodologies}{Mixed / augmented reality}{}{}
  \CCScatTwelve{Computing methodologies}{Virtual reality}{}{}
  \CCScatTwelve{Computing methodologies}{Perception}{}{}
}

\begin{document}


\firstsection{Introduction}

\maketitle

\begin{figure*}[t]
\centering
  \includegraphics[width=1.0\textwidth]{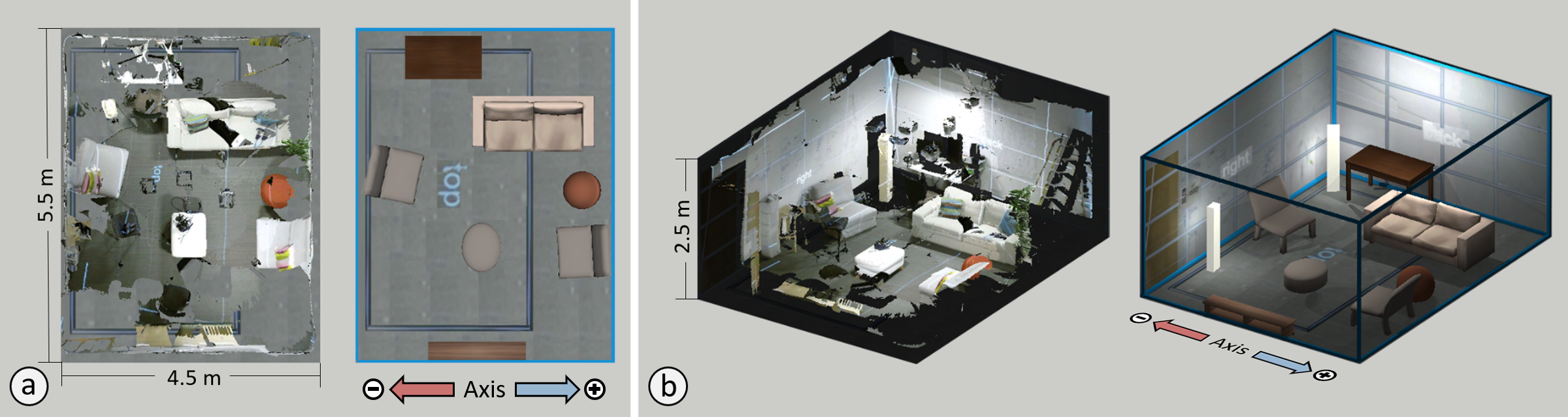}
\caption{a. Top view of the 4.5 m x 5.5 m room with furniture where the study and distortion effects were conducted. Left: Floor layout as scanned by Kinect v2 sensor cameras.  Right: Digital Double 3D model of room. b.	Left: Using SLAM (Simultaneous Localization and Mapping), a live 3D reconstruction of the room from a side angle. Right: Digital Double 3D model of the room from the same angle. 
}
\label{fig:pic3}
\end{figure*}

\begin{figure}[t]
\centering
\includegraphics[width=1\columnwidth, height=6cm, keepaspectratio]{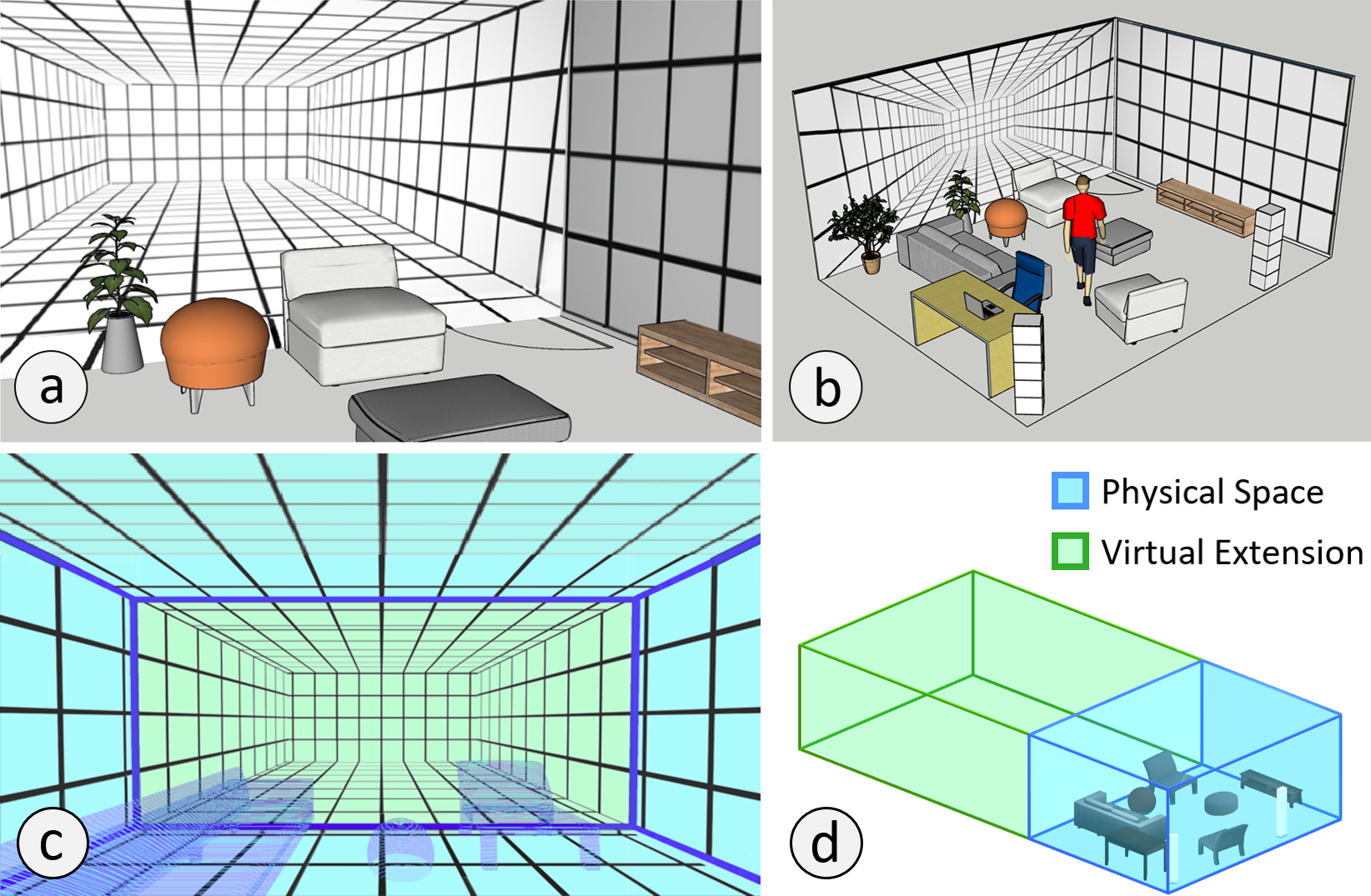}
\caption{a. First-person point of view (POV) from the perspective of a user viewing a virtually extended space in the room. b. Overhead view of the room environment where the user experiences the distortion effects. c. First-person POV as seen via head-tracking and perspective correction, where the green area designates the space that is extended through projection using the Elongation Distortion. d. Overhead view of the physical space of the room compared to its virtual extension during the Elongation Distortion.}
\label{fig:pic2}
\end{figure}

A growing number of virtual experiences take the user’s physical environment into account, which leads to an expansion of potential and possibilities in immersive home entertainment. Many gamers consuming entertainment in their homes are increasingly turning to immersive experience technology such as virtual reality (VR) and augmented reality (AR) for an extended reality or presence platform experience~\cite{home2022schell, world2022oculus}.

Some VR work specifically addresses the question of supporting navigation in large VR environments while relying on real walking in smaller physical environments. Redirected walking offers natural locomotion with correct proprioceptive, kinesthetic, and vestibular stimulation, but it requires sizable actual tracking spaces~\cite{nilsson201815}. Interaction-based redirected walking uses techniques such as warping~\cite{dong2017smooth, williams2021arc} and sensory technologies~\cite{sun2018virtual} perceptual illusion~\cite{sra2018vmotion}, or space mapping~\cite{sun2016mapping, hartmann2019realitycheck}. These techniques can operate in a smaller space; however, the experience is regularly interrupted to correct the user's position when the user approaches the limit of the available walking space.

VR routinely utilizes techniques such as room marking systems to assist users in navigating safely within the room when using VR. Room setup features introduced in SteamVR and Meta Quest SDK allow users to mark out surfaces of the physical layout of the home to better avoid collisions when the user is immersed in virtual reality~\cite{steam2022valve, sdk2022oculus}. These methods, such as collision bounds and Chaperone, cannot be directly applied to AR applications as the physical layout is present at all times in the platform experience. While redirected walking in Mixed Reality using VR headset and passive haptics~\cite{kohli2005combining, suma2013redirected} has been examined, our research marks a step towards redirected walking in visual AR. 

Motion parallax and perspective-correct rendering of computer graphics content allow augmented experiences that are different from the physical layout the viewer is in~\cite{gibson1979theory, bruder2012tuning}. In this work, we additionally explore the possibility of subtly manipulating a user’s natural locomotion via visual motion effects.

Reality Distortion Room (RDR) presents five room distortion treatments in augmented reality that employ the user's visual perception and spatial understanding to subtly manipulate their position via natural locomotion. Using wireframe overlay effects, we mapped walls and furniture to generate an omnidirectional room-scale display that renders a 3D reconstruction and extension of the user's physical space. Out of the five distortion treatments, three treatments are designed to impact the user's directional motion pattern (movement along an axis, refer to Fig. \ref{fig:pic3}a for axis directions), whereas the other two treatments are designed to impact the user's motion to and from the room center (distance-to-center). 

Our study examines natural locomotion and visual perception. We selected RealityShader to simulate distortion treatments because it seamlessly blended projected and physical environments while the physical layout remained undistorted~\cite{youtube2021realityshader}. AR headsets with see-through waveguide displays, such as HoloLens 2 or Magic Leap 2, are equipped with relatively small field-of-view displays, which limits immersion. Video pass-through MR, such as the Varjo XR-3 or Meta Quest Pro, is becoming more commonplace but still has fidelity problems. Thus, we opted to run this study in spatial augmented reality, using spatial projection as seen in Fig.\ref{fig:teaser}a. We utilized RealityShader's projected augmented reality system~\cite{jones2014roomalive} to capture and react to the user’s locomotion response. The RealityShader system allows us to assess visual distortion effects while enabling users to walk in a fully surrounded projected space and untethered to a physical device as seen in Fig. \ref{fig:pic2}b. We performed several additional treatments, including augmenting the space with randomly floating particles and overlaying furniture outlines as seen in Fig. \ref{fig:teaser}.

Our system encourages users to move in certain ways within the room. In applying the distortion treatments in an immersive AR experience, our study found that the Reality Distortion Room impacts users’ movement and reactions in specific ways, without explicitly telling them to do so. This study provides the first empirical evidence that developers can influence the motion of users by warping and modulating the AR space, which suggests a potential mechanism to be used in the eventual realization of redirected walking in an AR environment. We see great application in our system for helping users make modest positioning adjustments, especially when standard tools for adjustments, such as sound systems, are being used. We make the following contributions: 

\begin{itemize} 
    \item Design and pilot testing of different distortion geometries for our system, the Reality Distortion Room.
    \item A user study (n=20) demonstrating the effectiveness of systematically influencing a user's natural locomotion. 
    \item  Analysis of study results, demonstrating user locomotion responses to generic room shape changes in a projected augmented environment. In particular, a directional effect and a center distancing effect are demonstrated as a reaction to the geometric deformation of the environment (distortion treatment) alone.

\end{itemize}

\section{Related Work}

Our platform offers a virtual world that alters and extends existing physical space. Making use of physical layout to enhance the experience in a virtual world was previously explored. Fuchs envisioned the potential ways to utilize the combination of virtual and real worlds, implementing the CAVE system in the Office of the Future project~\cite{raskar1998office}. 
The spectrum of the AR-VR continuum was previously explored from the human computer interaction aspect~\cite{milgram1994taxonomy, sayyad2020walking, benko2014dyadic, benko2015fovear, shapira2016reality} Each method utilizes space in different ways when users are exploring virtual environments. For example, one such method detaches the user entirely from the physical space, making the movement frictionless and stationary using a treadmill~\cite{darken1997omnidirectional,iwata2005circulafloor,iwata2006powered,median2008virtusphere}. Other methods include adjusting sensitivity of input and output of the tracked movement between virtual and real to redirect the user~\cite{langbehn2018redirected, sun2016mapping} or providing a dynamic haptic environment where the physical layout adopts to a virtual one~\cite{iwata2005circulafloor, cheng2015turkdeck, suzuki2020roomshift}. In addition, previous work investigating ``vection'', or ``illusions of self-motion'' study how to convincingly simulate human locomotion in virtual environments without having to allow for full physical movement of the user~\cite{riecke2012selfmotion, riecke2013perceptual}. RDR, on the other hand, looks at the problem from the perspective of inducing the full physical movement of the user without explicit guidance systems or instructions. Therefore, we designed a system to induce movement patterns in user locomotion without informing them about the desired movement pattern, meaning participants were only instructed to move around freely in the environment. Inspired by these earlier studies, our work embraces the situated physical reality and our senses within it in our mixed reality experience. Through our proof of concept and user study, we demonstrate how to subtly manipulate user locomotion in AR space. 

\subsection{Experiencing Large Space}
Experiencing a larger environment than the one in which one exists can be both physically and conceptually disengaging for users~\cite{hartmann2019realitycheck}. To overcome these issues of disengagement, creative measures are taken to redirect attention and enable redirected walking~\cite{razzaque2001redirected, sra2018vmotion, sun2016mapping, peck2010improved} through procedurally generated virtual space from 3D reconstructed physical space~\cite{sra2018oasis, sayyad2020walking}. Event-based methods, such as dynamic saccadic redirection, have showcased a way to reduce the space required for immersive experiences~\cite{sun2018virtual}. Remixed Reality~\cite{lindlbauer2018remixed} explores the direct manipulation of the environment by offering users different and larger room layout options from their perspective. To further extend virtual space through augmentation and remote user presence, Room2Room~\cite{pejsa2016room2room} presented a prototype implementation for rearranging and extending the virtual layout in consideration of physical space through augmentation. Additionally, Reality Check~\cite{hartmann2019realitycheck} demonstrated how a virtual game environment can be combined with real-time 3D reconstruction of the room, resulting in a presence platform experience.

\subsection{Room-scale Interaction}
The physical constraints of a room can make it difficult to fully experience the six degrees of freedom in virtual reality. However, extensive research led to the exploration of redirected walking~\cite{razzaque2001redirected, nilsson201815, steinicke2010estimation}, a method that subtly adjusts the user's direction without them noticing, allowing the user to be immersed in a large virtual environment within limited physical layout they are operating in to provide a seamless experience~\cite{williams2021arc, wilson2018object, langbehn2018evaluation}.

Recent works~\cite{jones2013illumiroom, jones2014roomalive, benko2014dyadic} adopted a spatial augmented reality (SAR) system, embracing limited room space and scale and delivering a customized layout. This platform suggests strategies that use a room for an access point, port, and physical interactive space while expanding the interactivity well beyond the room scale.

In recent years, we have seen many room-scale VR games that use the layout of the room both as part of the virtual environment layout and use the full floor space as an interactive space~\cite{rune2020eye, curiousvr2022custom, tea2022void}. For example, the VR game Custom Home Mapper: Castle Defender (2020) uses the player’s room layout to generate the top floor of the tower balcony. Eye of the Temple (2020) constructs a temple maze from surroundings for the player, and Tea for God (2021) converts the player’s room to a bunker with windows on each wall. Room-scale interactability is appealing as it can be catered to current VR users, which is why research targeting room-scale mixed reality persists~\cite{langbehn2018evaluation, sra2018vmotion, gal2014flare}. Our system also caters to the standard room-scale layout as seen in Fig. \ref{fig:pic3}.

\begin{figure}[t]
\centering
\includegraphics[width=0.8\columnwidth, height=4.0cm, keepaspectratio]{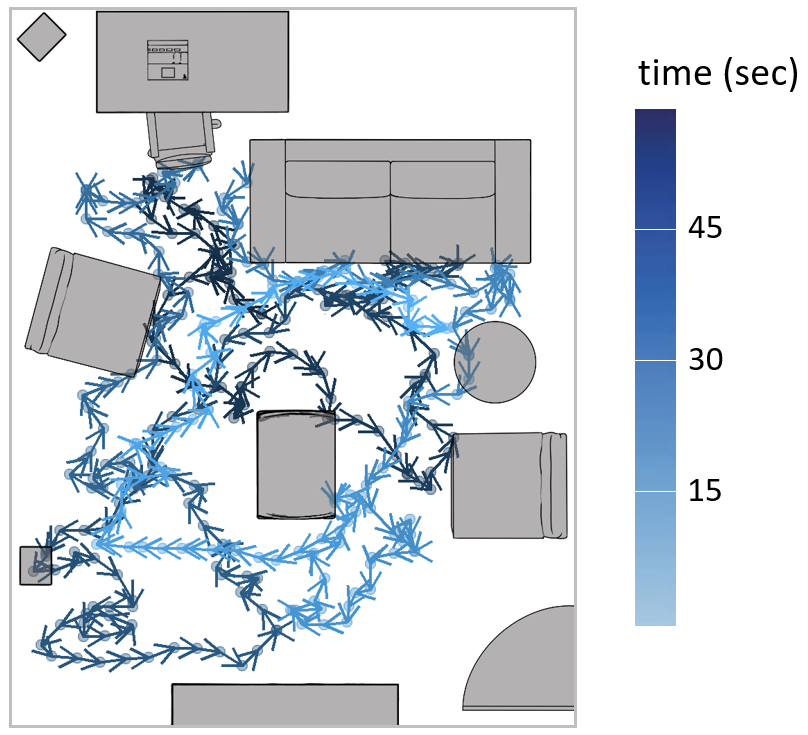}
\caption{Example path of a user moving around the room during a one-minute trial, demonstrating full possible utilization of the physical space without fear of bumping into objects in the dark environment.}
\label{fig:pic33}
\end{figure}

\subsection{3D Reconstruction for Physical Spaces}
The wide accessibility of depth cameras has produced much remarkable research around 3D reconstruction using these technologies~\cite{newcombe2011kinectfusion, peasley2013replacing} such as adding IMU sensors~\cite{giancola2018integration} to examine outdoor 3D navigation~\cite{qayyum2013kinect}. Robust 3D reconstruction in a combination of object detection research~\cite{lai2013rgbd, alexandre20123d, li2020volumetric, meka2020deep} provides insight into always-on display AR experiences~\cite{nassani2015tag, ibrahim2018arbis}. HoloLens and Magic Leap, both augmented reality (AR) devices, utilize spatial mapping~\cite{magic2022what, zeller2022spatial}, scan and examine the physical layout of the user's surroundings to find the optimal location to place virtual contents~\cite{kumaran2023impact, kim2022investigating}. Projects like FLARE, SnapToReality, VRoamer and Dynamic Theater exhibit the potential these intelligent virtual content projections have to enrich immersive mixed reality experiences~\cite{gal2014flare, nuernberger2016snaptoreality, cheng2019vroamer, kim2023dynamic}. While current spatial mapping for VR-AR systems does well with scanning surroundings for planar surfaces, finding usable space to project virtual objects, and designing a virtual environment, they are ultimately restricted to the physical layout. 

The benefits and impacts of a wide field of view beyond our vision are clear~\cite{ren2016evaluating}. Our projection mapping system utilizes a 3D reconstructed model of the room to create a view from the virtual environment. That view is then projected onto real physical surfaces such as the walls and furniture in the room, simulating the perspective of a co-located virtual world~\cite{hartmann2019realitycheck} using the real world as a baseline.
We conduct our user study in a full-surround spatial augmented reality system, augmenting the entire human field of view and beyond, in order to create the most convincing user experience of the transformation of their surroundings.

We designed a set of virtual environments that are aligned or partially aligned with the physical room that is deformed, extended, and subtracted from the perspective of the user's eyes as demonstrated in Remixed Reality~\cite{lindlbauer2018remixed}.
In our system, the user sees the physical world at all times as seen in Fig. \ref{fig:teaser}b. We build the experience around the room layout including the furniture in the room. 

\section{Reality Distortion Room}
Reality Distortion Room uses the RoomAlive infrastructure~\cite{benko2014dyadic, jones2014roomalive} to deliver a full surrounded augmented reality experience. The room geometry is scanned and loaded into the game environment and Unity workspace where sets of geometric space transformation (distortion treatment) are deployed. Virtual models are placed into the scene in relation to the physical room to project reconstructed geometry. This allows our system to extend the virtual world from the real world, as if the room is transforming, by rendering the physical world within the scene. To simulate the changing geometric environment through the projected walls of the physical room, the synchronization between the user's head position, digital twin, and the real world is crucial. The live reconstruction of the physical environment is done using four RGB-D cameras (Microsoft Kinect v2) that are placed in each corner of the room. The real-time geometric representation of the world is then directly placed in the virtual game environment that warps and enlarges while tracking the head position to reflect the user's perspective.

\subsection{System Infrastructure} \label{sec:system}
The AR projected room of approximately 4.5 × 5.5 meters incorporates four Microsoft Kinect v2 depth cameras in each corner of the ceiling line. We placed the furniture objects as we would in our own living room, making a close representation of the living room where the user would use our system. Five wide field-of-view projectors render a 360$^{\circ}$ Spatial AR system, projecting to surfaces within the room and fully utilizing the objects inside the room including the furniture and moving objects. Virtual objects are presented from the user's viewpoint as found in RoomAlive system~\cite{jones2014roomalive}. The distorting 3D geometry model is placed and oriented in the physical room and the scene is constructed based on the viewpoint of the user. As the projection is rendered in a view-dependent manner, the participant is free to walk around the room without a headset. As the user in the room is not tethered to anything they are encouraged to walk around and conduct true normal locomotion. Since Reality Distortion Room features full surround visual coverage of the virtual environment, projected onto the physical environment, what the user will see is the room they are standing in, transformed into a different shape. While the distortion treatment is underway, the user's stereopsis will not align with the intended visual of a deformed room. We made sure that all our distortion treatments return back to the default physical room layout with the standard projection mapping applied. 
We believe coming back to a condition where the virtual environment aligns with the physical arrangement is crucial for the user's affordance as this process blends real and virtual. For the virtual room to be precisely aligned with the physical counterpart, the system goes through a calibration process. Static 3D geometry that includes both stationary and moving objects is reconstructed with data collected from scanning a cloud of 3D points. With these baseline dimensions, projected content may be precisely aligned with the physical layout. Details on this calibration process can be found in the RoomAlive paper~\cite{jones2014roomalive}.

\subsection{Distortion Treatment Components}
Distortion treatment alters the geometric perception of the physical layout. We implement various transformations of the environment. We evaluated various distortion treatments to determine their effectiveness in generating consistent locomotion transitions. We examined 10 treatment designs: elongation, warp, shift, elevation, enlarge, enlarge (even larger), rotation, twist, furniture rotation, and furniture shift. Highlighted grid panels, where each tile is 65cm x 65cm, overlay the entire room assisting the user in understanding the transforming geometry while showing the scale transformation. We also added particles in the space to demonstrate accurate reflections of motion parallax as users moved as well as to assist with seeing added or subtracted space. This easily allows better spatial awareness, while inducing more movement without presenting one target. Each particle is 1.92cm in radius and floats at a speed of 1 cm per second in a random direction (Fig. \ref{fig:teaser}c). Approximately 712 particles float around in every 10-meter cubed space. Every trial we tested featured particles in space except for one trial, which was baseline without distortion.

While designing three distortion treatments that stimulate the occupant's movement along an axis identified in Fig. \ref{fig:pic3}), we imagined how we would move and respond to a transforming space around us. We identified where people would be most likely to walk toward to secure their safety or view when the room began distorting. The same concept was applied to the two distortion treatments that influenced the user’s distance to the center of the room. When designing the floor layout of the distortion treatment we made sure the virtual space did not transform any smaller than the physical walkable space, presenting the full room layout for users to walk around without being able to step out from the virtually rendered space. 

\begin{figure}[t]
\centering
\includegraphics[width=1\columnwidth, height=6cm, keepaspectratio]{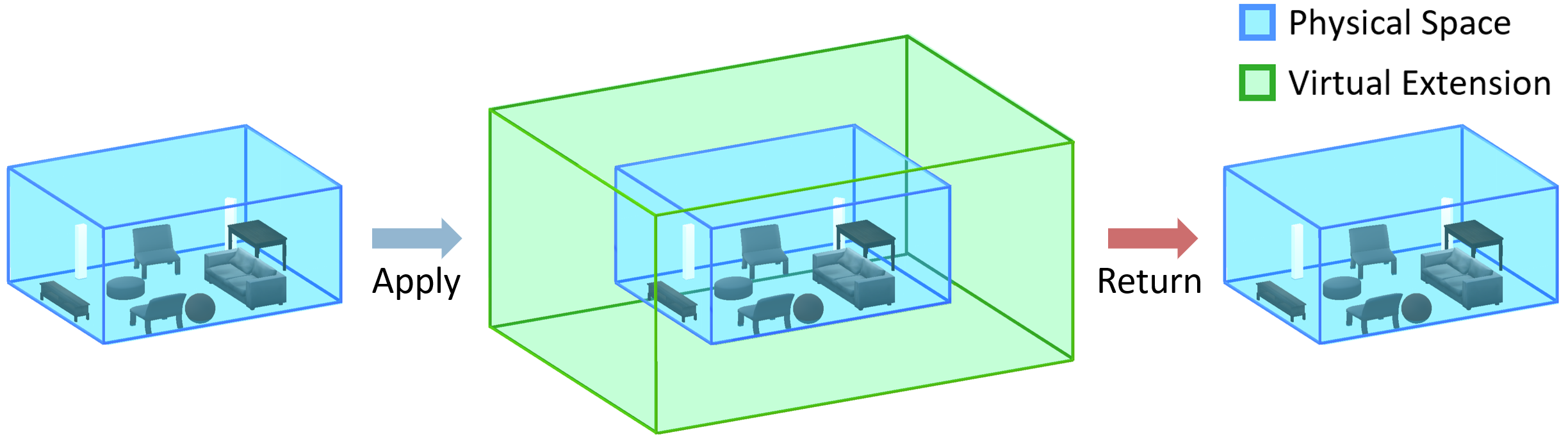}
\caption{The room transformation process during active distortion treatment, using the Expansion Distortion as shown above. During the ``Apply'' segments, the virtual room enlarges for 10 seconds. This is immediately followed by the ``Return'' segments, where the virtual room is compressed for 10 seconds until the virtually extended space merges back to the original room layout.}
\label{fig:pic6}
\end{figure}

Each distortion treatment is divided into two phases: apply and return. Apply represents the first segment where the distortion treatment begins, and return entails the latter segment where the virtually distorted room reverts to the original physical room layout. The timeline of the transformations used in each trial for all distortion effects can be seen in Fig. \ref{fig:pic6}. 

The dependent variable is the participant's reactions, which we measured through user locomotion. User locomotion is when a participant moves around the room while an assigned stimulus is being applied or returned. The Reality Distortion stimulus was measured as either a directional effect (user movement along an axis) or the Central Effect (change of the user's distance to the room center) and depends on the design of the distortion treatment.

The directional effect measures the user’s movement in relation to the axis along the room's smaller dimension (i.e. left and right in Figure \ref{fig:pic11}) while the apply and return stimuli are in effect and consists of Elongation, Warp, and Shift Distortions. The central effect (distance to center) measures the change in the user’s positioning  away from or towards the center of the room compared to before/after the stimulus segment and consists of Elevation and Enlarge Distortions. The average of the total user movement during each stimulus segment (10 seconds) was used for our evaluation.

Baseline refers to the situation in which no distortion treatment is augmented. We established a baseline by running trials with particles but no visual effects or treatments. We separated the baseline data into 10-second segments for the analysis. We randomly selected 15 segments from each user's trial to include all potential data segments from the trial.  Along with the five distortion treatments, we conducted two additional trials: 1) no distortion treatment with no particles and 2) no distortion treatment with particles. As all trials with distortion treatment included particles, base data points were collected in trials with particles although no distortion treatment was applied.

\begin{figure*}[t]
\centering
  \includegraphics[width=0.85\textwidth]{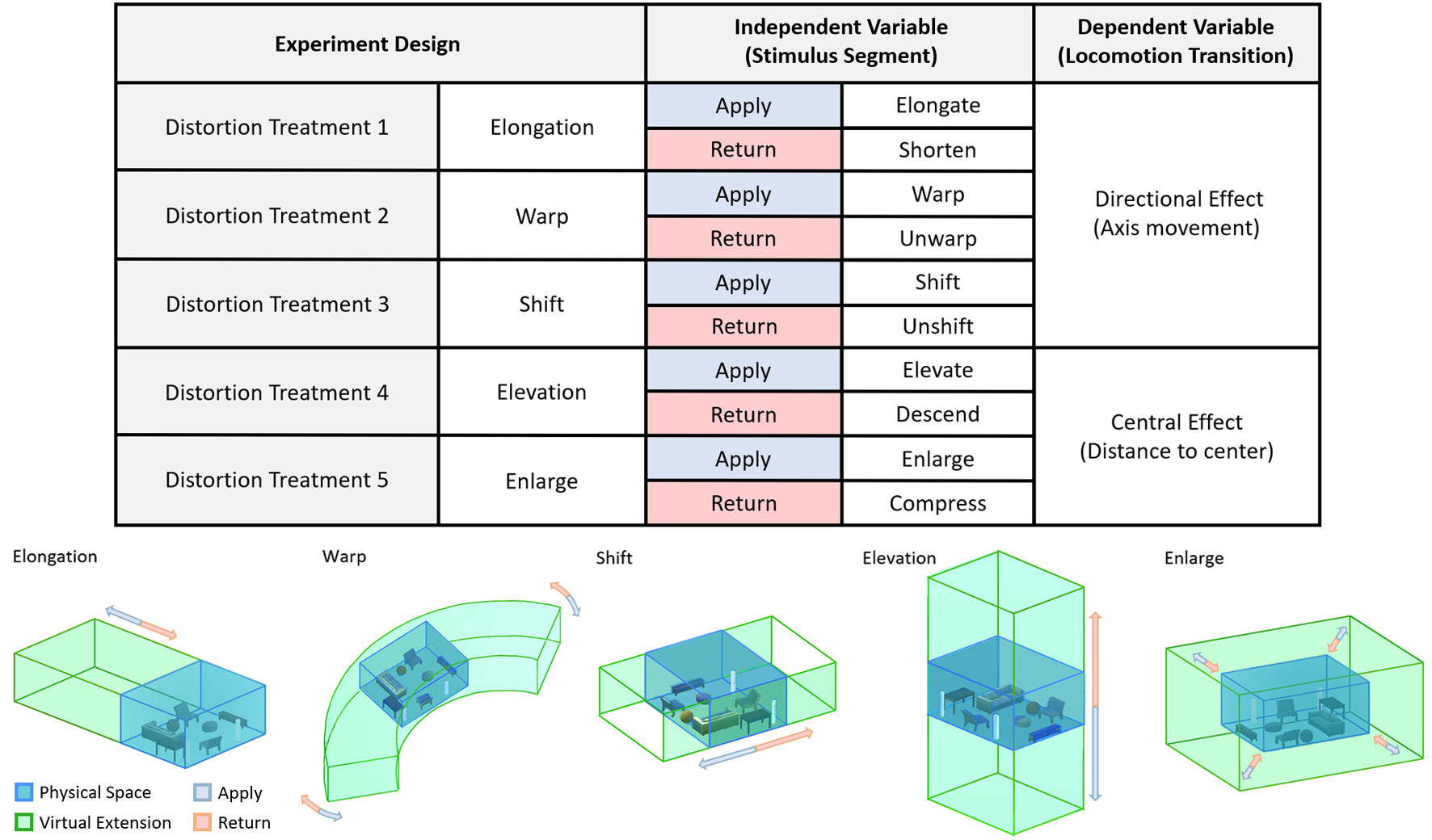}
\caption{Distortion treatments 1, 2 and 3 stimulate axis movement, while distortion treatments 4 and 5 stimulate distance to the center.}
\label{fig:pic5}

\end{figure*}

\subsection{Pilot Study}
The experimental portion of this study was designed to measure the effect of distortion treatment on user locomotion responses. First, experiment design (distortion treatment) is a set of geometric distortions that affect the user's movement patterns within the room along the axes, central point, and rotation. All the distortion effects occur with respect to the physical room, regardless of the orientation of the user. We examined all ten distortion designs in our pilot study. In addition, each treatment was tested with two different segment speeds (7 seconds and 10 seconds), for a total of twenty treatments in which ten induced the directional effect while the other ten induced the central effect. 

We also assessed the impact of deploying particles to encourage increased walking. We found that distortion effects characterized by excessively rapid transformations, intricate geometric alterations, and rotation effects did not conform to discernible patterns. For example, some participants noted difficulty comprehending the transformations based on the speed or geometry of their implementation, citing dizziness and confusion. Based on this feedback, We omitted certain distortion treatments, altered the geometry and speed of the remaining transformations, and here are the five distortion designs included in the main study:

\begin{itemize} 
    
    \item Distortion treatment 1 (Elongation) refers to a geometric layout transformation of a room where one wall recedes into the distance and then returns (elongates and shortens). This is called Elongation Distortion and the treatment is designed to have a directional effect. The Elongation Distortion (Fig. \ref{fig:pic2}) elongates one side of the wall outward horizontally for 3.35 meters during the ``Apply'' segments of the stimulus, while in the ``Return'' segments the elongated wall shortens back to align with the physical room. 
    \item Distortion treatment 2 (Warp) refers to a geometric layout transformation of a room where the entire room warps and unwarps. This is called the Warp Distortion and the treatment is designed to have a directional effect. The Warp Distortion consisted of the continuous warping (bending) of extended versions of two opposite parallel walls of the room. The distortion treatment consists of one warp and unwarp action for each stimulus segment. The bend angle of the parallel walls consists of 160° with the bend executed along a 19.33m length of the wall. 
    \item Distortion treatment 3 (Shift) refers to a geometric layout transformation of a room where parallel walls shift and unshift (shift back) horizontally. This is called the Shift Distortion and the treatment is designed to have a directional effect. The Shift Distortion utilizes two parallel walls of the room, shifting 5.14 meters side to side in the horizontal direction. During the ``Apply'' segments the wall shifts in a set horizontal direction, while in the ``Return'' segments (unshift) the wall shifts back in the opposite horizontal direction.
    \item Distortion treatment 4 (Elevate) refers to a geometric layout transformation of a room that ascends from and descends to the ground level. The user experiences this as either being elevated above or sinking below the ground. 
    This is called the Elevation Distortion and the treatment is designed to have a central effect (changing the user's distance to the room center). The Elevation Distortion consists of the virtual room going up and down 8.07 meters vertically. During the stimulus segments, the room elevates from and descends to the base ground level.
    \item Distortion treatment 5 (Enlarge) refers to a geometric layout transformation of a room that enlarges and compresses. Users experience this as the walls moving away from or coming closer to them. 
    This is called the Expansion Distortion and, like for Elevate, the treatment is designed to have a central gathering or dispersion effect. The Enlarge Distortion consists of two segments: virtual room 'enlarge' and 'compress'. The room enlarges to double the width, length and height while expanding its volume from 56.82 m$^3$ to 454.56 m$^3$.
    
\end{itemize}

\section{Experiment}
We recruited 20 participants, ages 23 to 32, of which eight identified as male and twelve identified as female. Four participants had previously tried VR while only two previously experienced AR. On average, the study lasted approximately 30 minutes, including the time needed for the instruction, and consisted of seven trials and a 10-minute interview. We conducted one trial for each distortion treatment and two additional baseline trials: no treatment with particles and without particles. The RDR system described in \ref{sec:system} System Infrastructure is used in a room environment that is approximately 4.5m × 5.5m × 2.5m (Fig.\ref{fig:pic3}). 
Based on dominant results from prior work in Spatial AR that highlight creative ways to interact with extended reality, and the ways that participants responded to controlled and comprehensive space, we defined the following hypotheses: 
\begin{itemize} 

    \item H1: The augmented distortion treatment can induce participants to move more in some directions than others.
    \item H2: The augmented distortion treatment can induce participants to move closer to or away from the center of the room.
    \item H3: The augmented distortion treatment can induce participants to turn or move in a circular direction.


\end{itemize} 

\subsection{Procedure}
Participants were introduced to all five distortion treatments in each trial in addition to two trials for the baseline: static room with particles and without particles in the space. Each trial lasted a full 60 seconds, consisting of two cycles of stimulus segments: two ``Apply'' segments and two ``Return'' segments.

When participants arrived, a researcher guided them into the room equipped with the projection system. Upon entry, the room projectors remained off though ambient lighting allowed the user to see objects in the room.  The ambient lights remained on through the entire trial but were outshone by the lights from the projectors. 

Participants were given a brief introduction to our system and verbal instructions for the study:  they are free to move around and interact naturally within the room during the trial but asked to refrain from sitting down on any surfaces. We encouraged walking and examining the room during the trial and let them know that we would be asking a few questions following the trial. We also informed participants that they were free to leave the room at any time if they felt discomfort (i.e. sickness, fright). A researcher remained outside of the room to monitor participants through the 3D reconstructed live inspector.

After each 60-second trial, we turned off the projection and asked the participants to sit in the center seat of the couch. In between trials, we asked a few questions about their experience to ensure the participant felt okay and ready to continue. Before moving to the next trial, we asked if they could describe in a few words what they saw and experienced. After the final trial, participants were asked to reflect on their experience of each treatment and safety concern. Finally, each participant was compensated with a {\$}10 gift card.   

\begin{figure}[t]
\centering
\includegraphics[width=1\columnwidth, height=8cm, keepaspectratio]{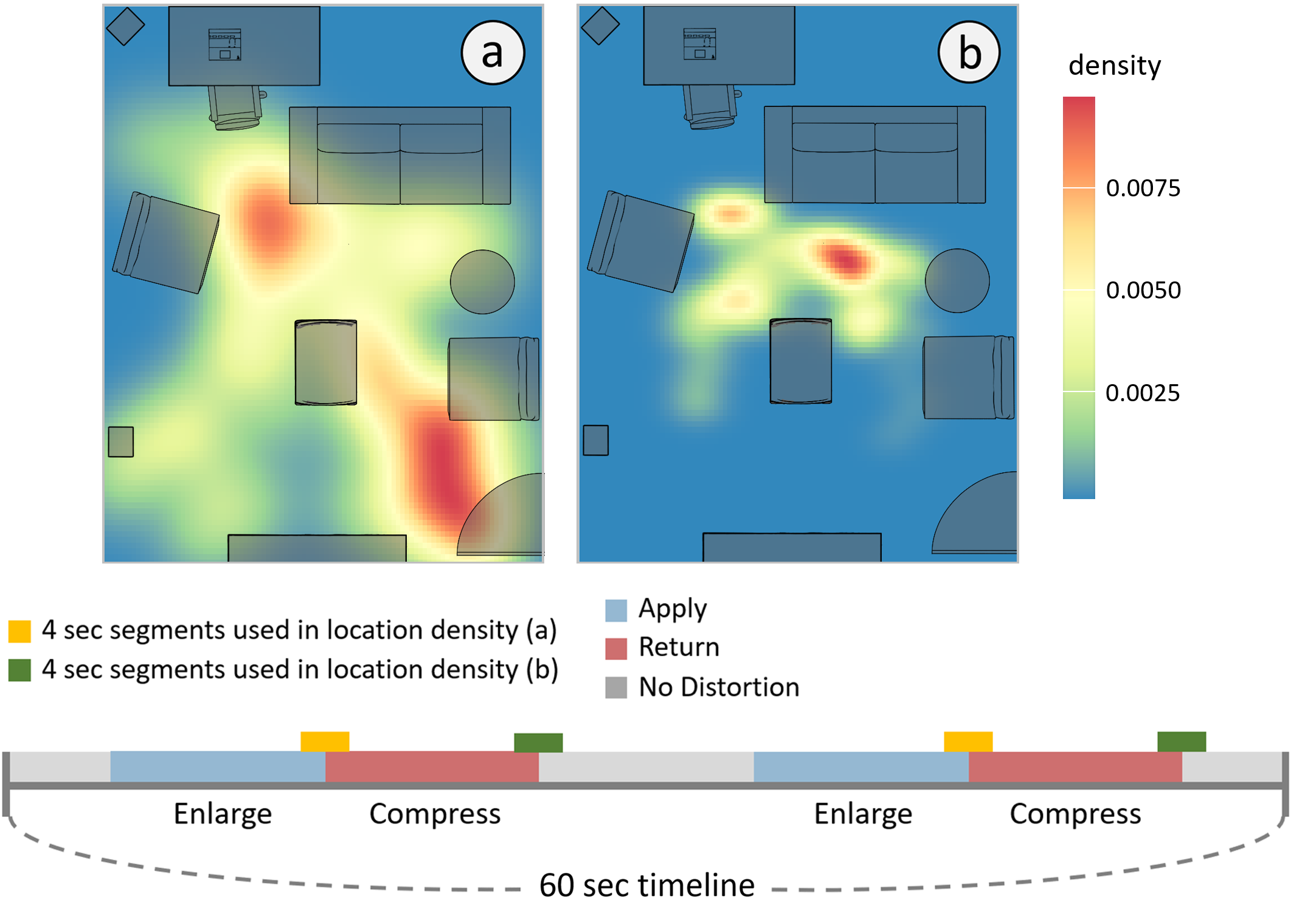}
\caption{The two location density maps, shown above, reflect data collected from twenty users during each segment: (a) Enlarge (apply phase) and (b) Compress (return phase). As shown in the timeline, collections are taken from two 4-second intervals: 2 seconds after and 2 seconds before the end of each phase of ``Apply'' and ``Return''. This shows how floor space was utilized immediately before and after the end of each phase.}
\label{fig:audioPerf}
 \end{figure}

\section{Results}
The experimental design (Fig. \ref{fig:pic5}) of this study intended to evaluate the changes in users' positions before and after the geometric transformation in the Spatial Augmented Reality room due to the application of distortion treatments. In this section, we report two types of Distortion Effects, with three distortion treatments designed for directional effect, and two distortion treatments designed for central effect. 

\subsection{Particle Effect and Natural Locomotion}
\begin{figure}[t]
\centering
  \includegraphics[width=1\columnwidth, height=4.5cm, keepaspectratio]{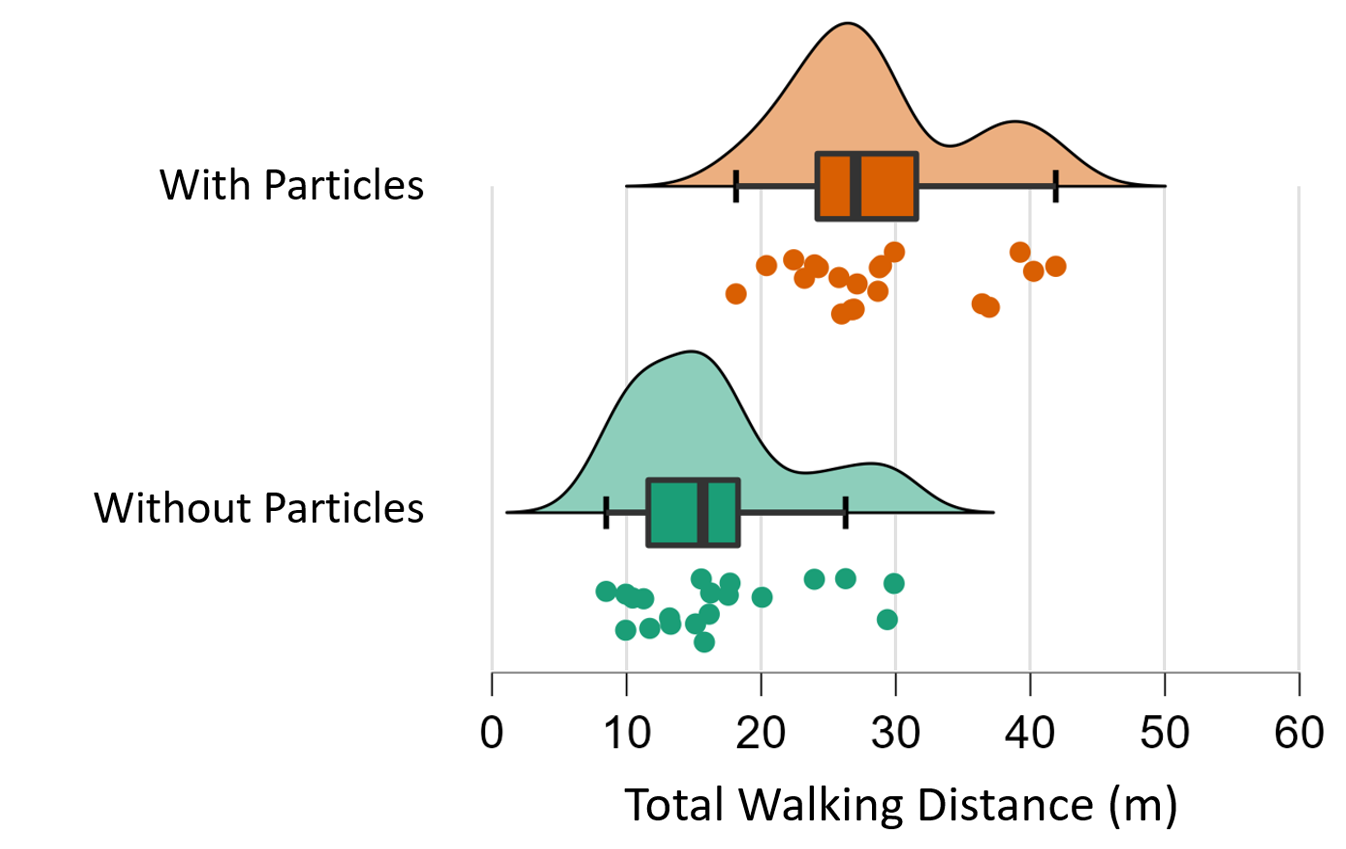}
\caption{Total walking distance (TWD) in effects with and without particles, revealing that the existence of particles increases walking distance ($p$ $<$ 0.001). The raincloud plot shows user distribution of TWD result and box indicating the median and interquartile range (IQR).
}
\label{fig:pic8}
\end{figure}

Our goal was to induce more natural walking without influencing the participant's movement in a specific direction, as there were no user tasks assigned. Here we compared the total walking distance in the room with particles, without particles, and without the distortion treatment. Without particles, the mean total walking distance was 16.70 m with a standard deviation of 6.38 m. Furniture outlines and wall grids were present in baseline trials with and without particles. Users walked an average distance of 28.80 meters, with a standard deviation of 6.75m, in studies with particles, revealing that the existence of particles significantly increased walking distance. ($p$ $<$ 0.001). As a result of an ANOVA analysis by classifying walking distance into a ``No particle'' group and a ``With particles'' group, there was a significant difference in the mean between the two groups (Fig. \ref{fig:pic8}c). In the ANOVA table, the $F$ value was 34.538, and the $p$-value was less than 0.001 demonstrating a statistically significant difference between the ``No particle'' group and the ``With particles'' group.

\begin{figure*}[t]
\centering
\includegraphics[width=0.8\textwidth]{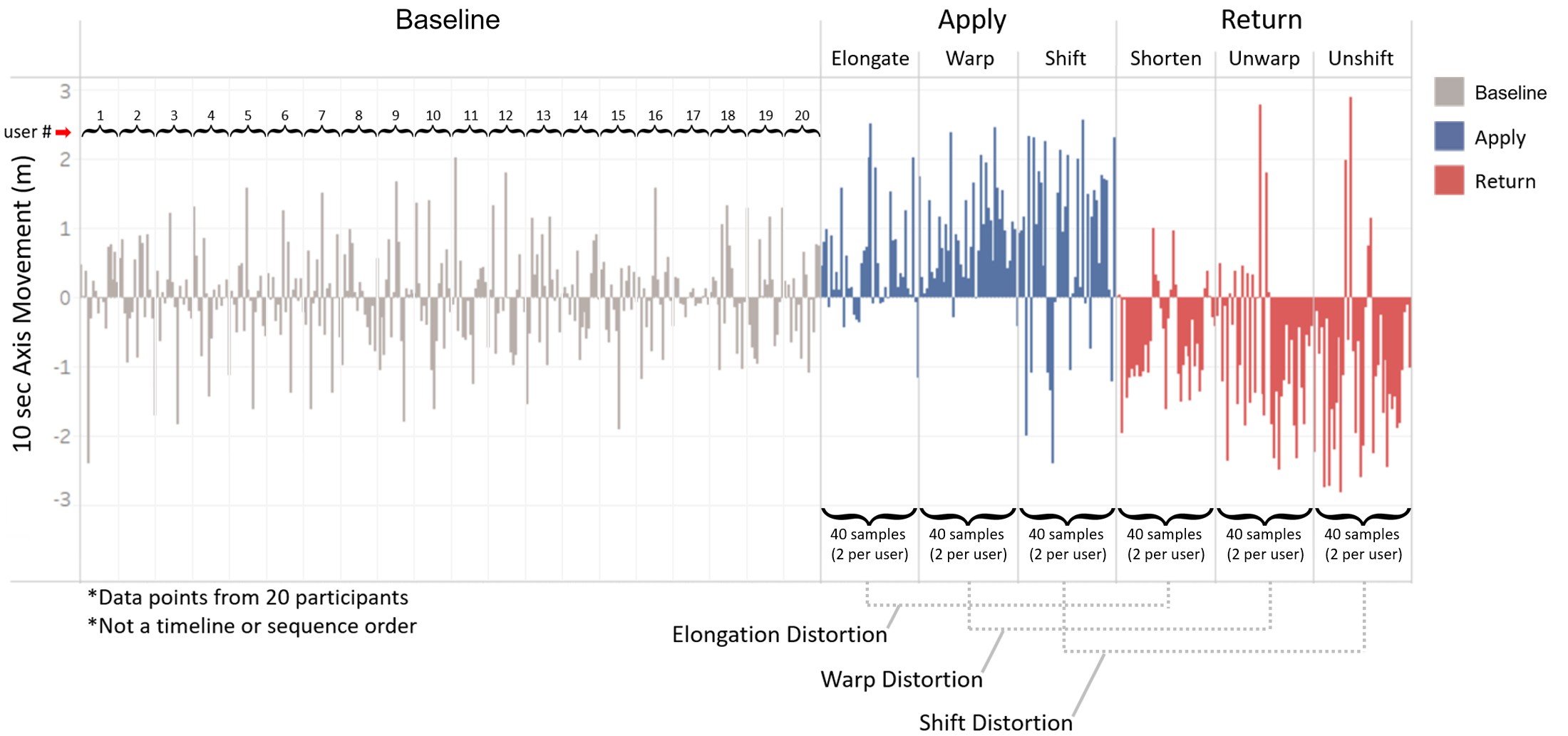}
\caption{Users’ movements on the axis of consideration corresponding to the effect applied to the room.
The chart shows data points (vertical lines) from each stimulus segment from 20 participants. Each vertical line shows data points denoting the average movement made in the axis direction over a 10-second period. We show stimulus segments among four conditions (Baseline, Elongation Distortion, Warp Distortion and Shift Distortion). Visible trends of users’ axis movement in the apply/return phase of three groups of distortion treatments become apparent by comparing with the baseline condition (see also statistical comparison in Fig. 10).
}
\label{fig:pic9}
\end{figure*}

\begin{figure}[t]
\centering
  \includegraphics[width=0.8\columnwidth, height=3.5cm, keepaspectratio]{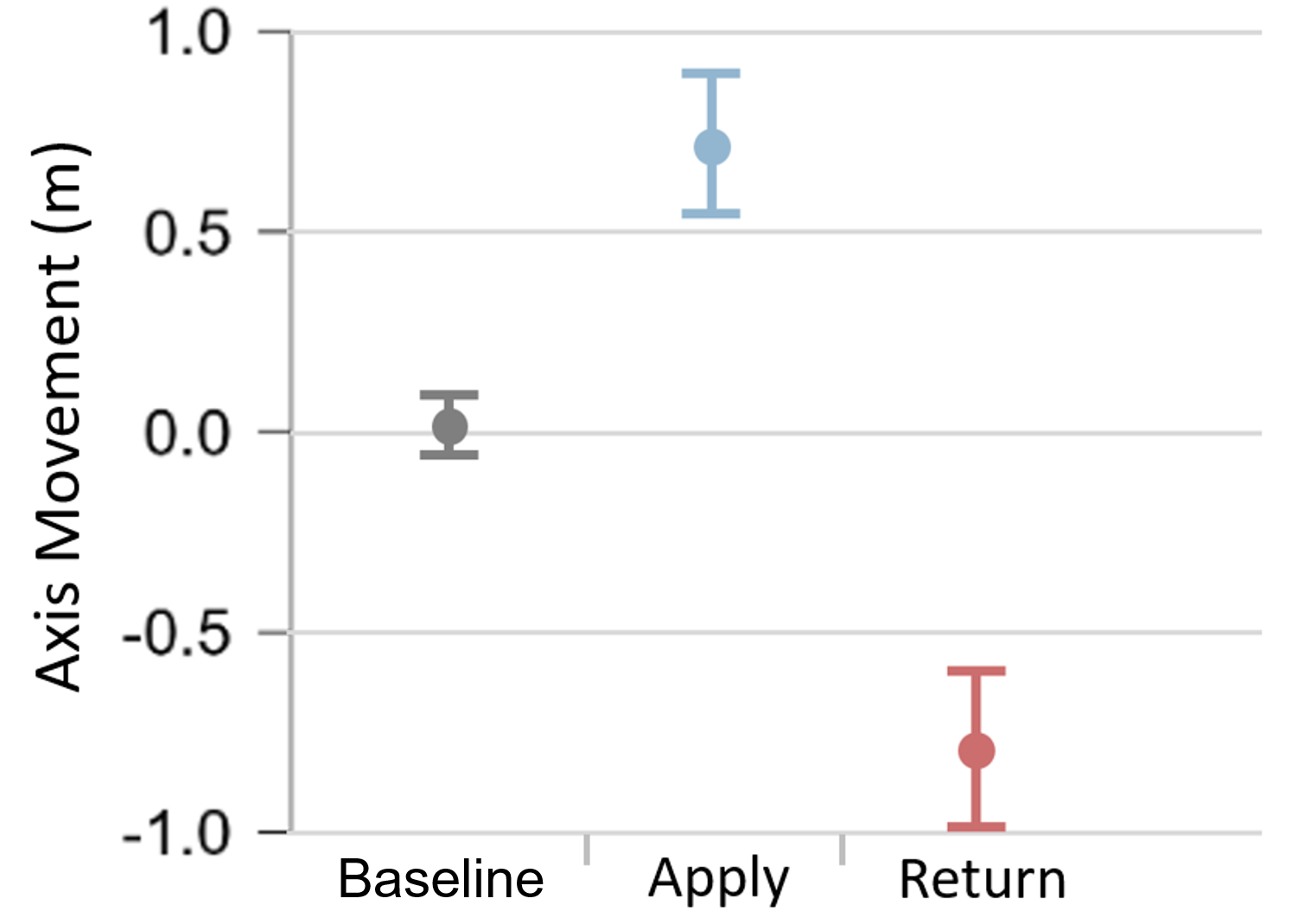}
\caption{Users’ movements on the axis of consideration corresponding to the effect applied to the room. ``Baseline'' represents the movement of twenty users during randomly chosen 10-second intervals without distortion effects. During the "Apply" segments, users generally moved in the positive axis direction and in the opposite direction during the "Return" segments. Error bars indicate 95\% \emph{CI}. An ANOVA with Bonferroni-corrected post-hoc pairwise comparisons reveals significant induced user movement during the distortion treatments.}

\label{fig:rotation}
\end{figure}

\begin{figure}[t]
\centering
  \includegraphics[width=0.8\columnwidth, height=8cm, keepaspectratio]{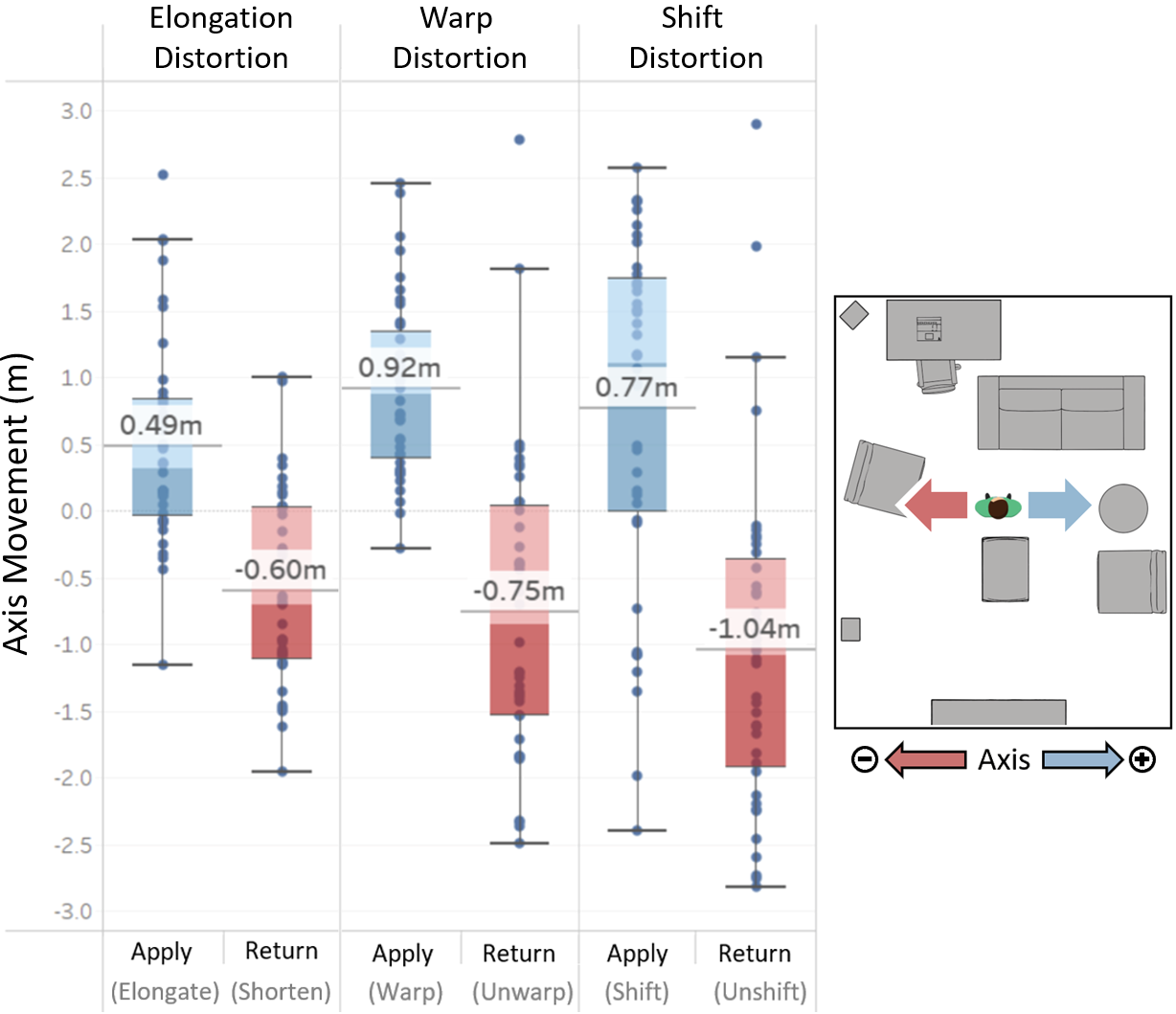}
\caption{Magnitudes of the apply and return of axis movements compared between three different distortion effects. Based on mean distance moved, the shift effect resulted in the most user movement towards the positive and negative ends of the axis, followed by the warp effect, and the elongation effect.}
\label{fig:pic11}
\end{figure} 

\begin{figure}[t]
\centering
  \includegraphics[width=0.8\columnwidth, height=3.5cm, keepaspectratio]{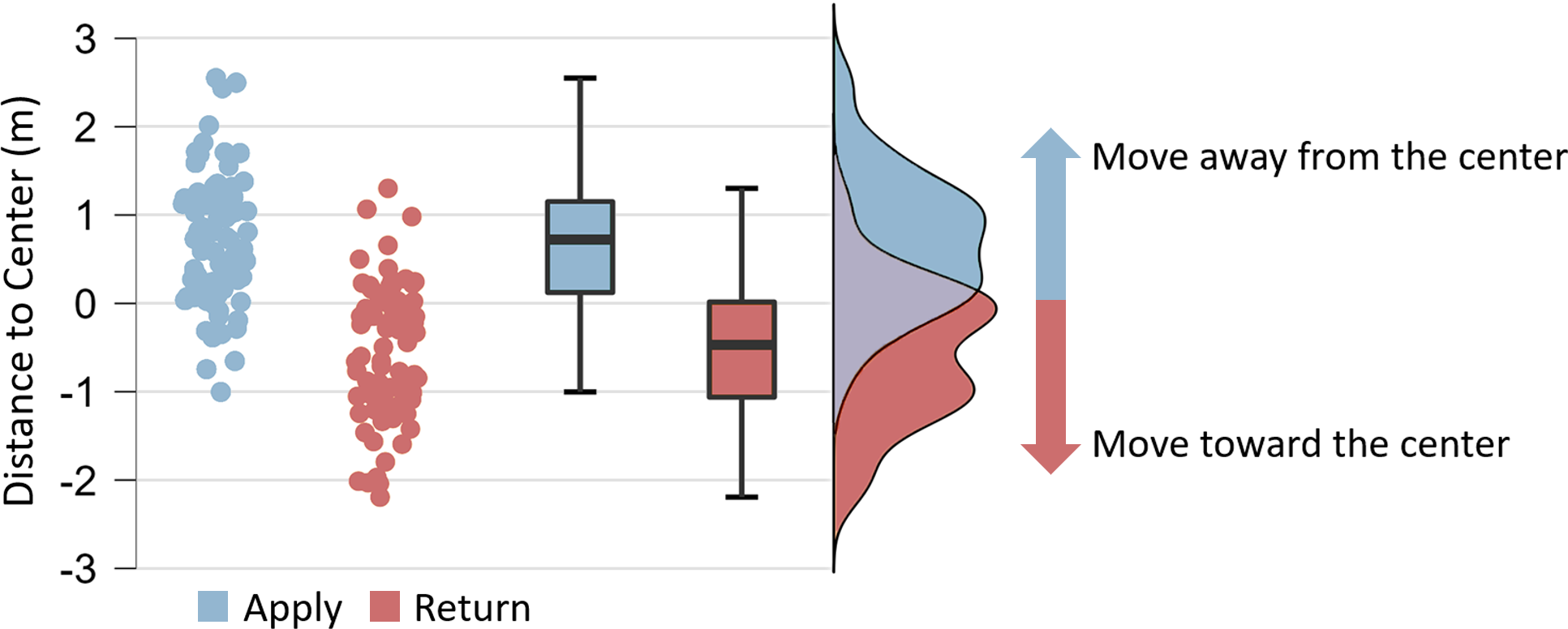}
\caption{A raincloud plot depicting movement away and towards the center of the room when the elevation and expansion effects are applied. The upper whisker boundary of the box-plot is the largest data point that is within the 1.5 IQR above the third quartile. According to apply and return, respectively, the distance  to the center was tracked.}
\label{fig:rotation}
\end{figure}

\begin{figure}[t]
\centering
  \includegraphics[width=0.8\columnwidth, height=8cm, keepaspectratio]{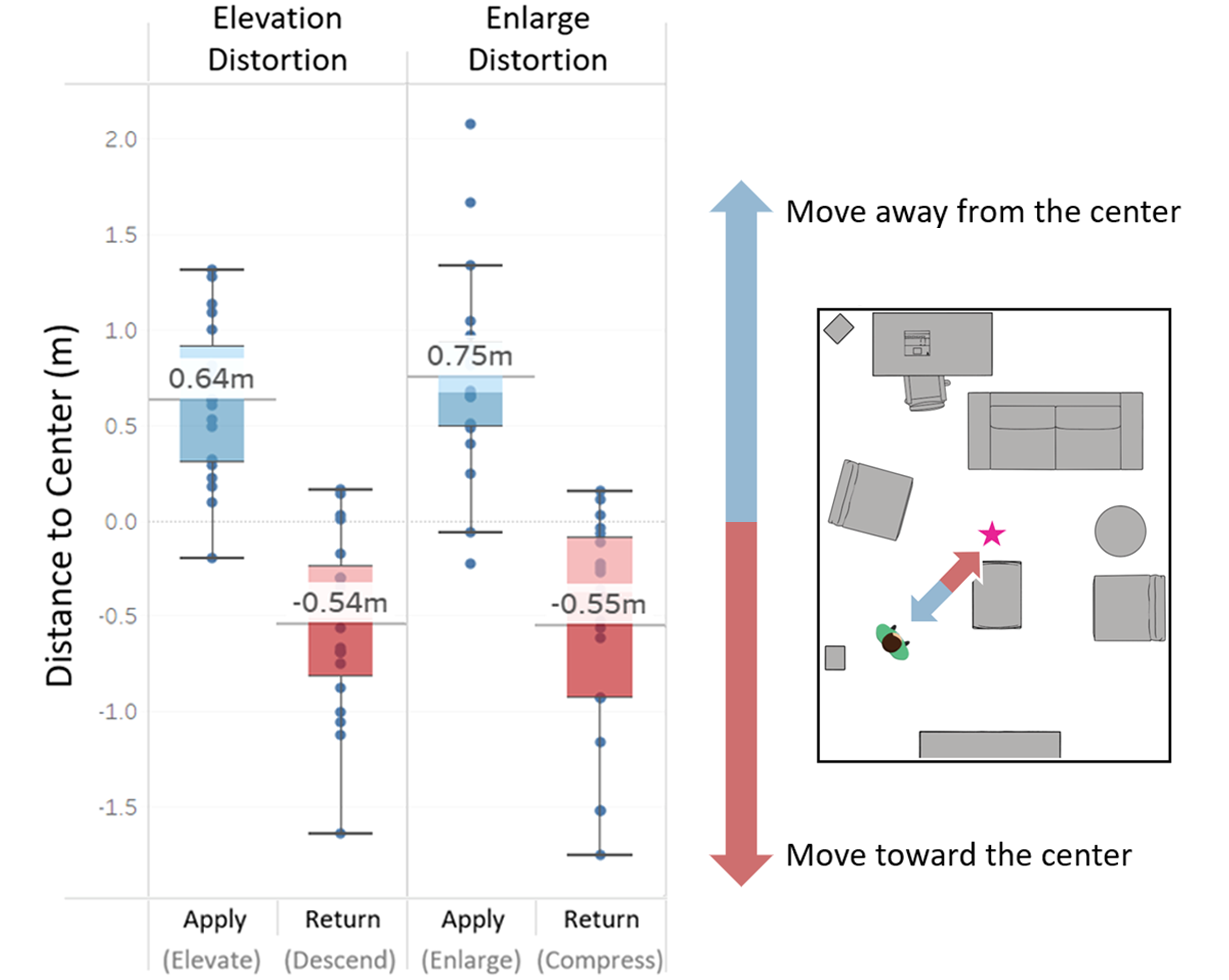}
\caption{Magnitudes of the apply and return movements measured how far a user was from the center of the room. Based on the mean distance moved, the expansion effect resulted in slightly more movement away and towards the center, followed by the Elevation Distortion.}
\label{fig:pic13}
\end{figure}

\subsection{Directional Effect: Axis Movement}
The following reports the directional effects of the three distortion treatments we designed and tested. We evaluated the 10-second user movement along axis direction within each stimulus segment. As seen from the data point of Fig.\ref{fig:pic9}, we saw a clear signal and trend between each stimulus segment and baseline. ``Baseline'' shows point data generated from 20 participants' baseline trials (no distortion treatment + particles) in two 10-second intervals throughout the trial. In the apply stimulus segment, users tend to move toward the positive axis direction (refer to Fig. \ref{fig:pic3} for axis directions), while moving in the opposite direction for the return stimulus segment. 

The experimental group was classified by Baseline Group, Apply Group, and Return Group and the ANOVA analysis showed significant differences between the three. In the ANOVA table between the three groups, the F value was 198.329, and the $p$-value $<$ 0.00005, 
showing a statistically significant difference in the size of apply segment and return segment. Post hoc analysis using Bonferroni adjustment also showed a significant difference in the mean between the three groups as follows. The difference between Baseline Group and Apply Group is $t$ value = -7.747, $p$-value $<$ 0.001. The difference between Baseline Group and Return Group is $t$ value=-8.924, $p$-value $<$ 0.01. The difference between Return Group and Apply Group is $t$ value=-13.948, $p$-value $<$ 0.001.

Each stimulus segment was tracked from apply and return, respectively. The movement of the axis for each group is analyzed as shown in the Fig. \ref{fig:pic11}. The average value of the return and apply segment is different for each group of the Elongation Distortion, Warp Distortion, and Shift Distortion. Based on mean distance moved, the Shift Distortion resulted in the largest directional user movement, followed by the Warp Distortion, and the Elongation Distortion.

\subsection{Central Effect: Distance to Center}
We analyzed the average movement change of the user throughout the 10-second distortion segments by comparing their position in the room before and after the segments. In other words, we measured the displacement of the user during the 10-second interval. Out of the five distortion treatments, two of them were designed and tested to manipulate the user's position relative to the distance from the center of the room. During the application stimulus segment, users generally moved away from the center, whereas during the return stimulus segment, users tended to move toward the center of the room. The remaining three distortion treatments had no consistent central effect on the user movement.

Experimental groups were divided into "apply" and "return" groups based on stimulus segments, and their distance to the center was analyzed using ANOVA. As a result, significant differences between groups are shown. In the ANOVA table, the $F$ value was 109.123 and the $p$-value $<$ 0.001, showing a statistically significant difference in the size of apply and return. The movement from the distance from the center for each group is analyzed as shown in Fig.\ref{fig:pic13}. Based on the mean distance moved from the center, the Enlarge Distortion resulted in the most user movement. This was found by comparing both stimulus segments although both distortion effects performed comparably. 

Overall, our study aimed to investigate the influence of augmented distortion treatment on users' natural locomotion while relying on their ability to comprehend the spatial transformation of the environment. We hypothesized that H1: The augmented distortion treatment can induce participants to move more in some directions than others and H2: The augmented distortion treatment can induce participants to move closer to or away from the center of the room.
    
Our user study results indicated that both H1 and H2 hold true. Additionally, the Shift Distortion showed the largest directional effect while the Enlarge Distortion showed the largest central gathering/dispersion effect. The third hypothesis (H3) aimed to test whether specific augmented distortion treatments could influence participants' natural locomotion to turn or move in a circular direction. During the pilot study, however, none of the designed treatments including those intended to induce circular movements, such as Rotation or Twist, were successful in producing this motion pattern. Participants also reported experiencing difficulty understanding and dizziness, which made it challenging to comprehend the nature of these transformations from Rotation or Twist Distortion effects. Therefore, the third hypothesis was not confirmed.


\section{Discussion}
As no clear objective was given, participants generally assumed that we wanted feedback on the space we designed, or that we were showcasing the new AR system. Many of them shared ideas on new 3D space ideas and what we should try next. This worked in our favor as we did not want to hint that we are examining their locomotion; our goal was to capture users' natural locomotion from the distortion treatment. 

We found compelling differences between the stimulus segments apply, return, and the baseline segment in locomotion response. Among all three distortion treatments for manipulating directional effect, the Shift Distortion had the largest distance manipulated before and after the stimulus segments, while the Warp Distortion treatment showed the biggest positive axis movement when the room is being warped. This is likely due to users trying to get a better vantage point to see the end of the hallway and get a better idea of what is happening in their surroundings. The Elongation Distortion had the weakest effect, but when comparing each stimulus segment to the baseline segment, the treatment still worked with the shortest mean average distance effect to axis. Of the two central effects, the Enlarged Distortion exhibited the most prominent distance-to-center effect before and after the stimulus segments.  

Only one participant removed themselves from the study, to answer a phone call. No participant requested a break during the distortion treatment augmentation or expressed sickness during our study. Participants' responses to Elevation Distortion showed a generally negative sentiment, stating that ``it took some time for me to understand what was happening'' (P4) and ``it was apparent when the room was going up but when the room was going down I wasn’t convinced'' (P17). In contrast, many participants expressed that the Enlarge Distortion was fun and refreshing: ``The room gradually expanding in all directions was my favorite'' (P13) and ``I felt like I was floating in space'' (P14). Two participants requested to experience the Enlargement Distortion again after the study. In contrast to these many positive experiences, a few participants bumped into objects during the enlarge stimulus segment. Lastly, most participants found particles in the space to be a nice addition, stating it ``magical''and ``fun to interact with'' (P5).

\subsection{Limitations}

Reality Distortion Room demonstrates a visual perception locomotion aid designed for the purpose of safety, entertainment, and interaction. While the study provides insight into how visual perception can be used to manipulate a user's natural locomotion, its findings are based on a small participant sample and brief trials, underscoring the need for deeper exploration of design considerations.

We tested five distortion treatment designs in this study understanding that many more distortion treatment designs could be explored. Our user study suggests interesting directions for future work and we foresee many ways to expand this concept. The challenge in using distortion treatment is that the transformation is very noticeable and, at times, intrusive. Creative solutions must be explored in using distortion treatment to deliver a cohesive user experience. We suggest distortion treatment to be adopted in catered space and situations, where the design of the geometric distortion is modified to the designated physical space. This curated experience, where aspects of the real-world deviate from reality, requires further study. 

Our study utilized a full-surround augmented reality platform with projections covering the entire human field of view and beyond, achieving this setup at home is unlikely. Additionally, while the system was the best fit for our use, whenever the user came within 60 cm of the wall, the user’s shadow became visible, blocking the projection of the wall. Also, the outline projected to furniture disappeared when the user blocked the projection. 

Though we had overwhelmingly positive responses from the participants, we believe the novelty factor may have played a role in how actively the participants explored the room, inducing more walking in the room. In addition, people moved without any instruction or clear objective in our study, as we wanted to examine if we could manipulate users without using instruction by only using distortion treatment. This raises questions about the performance of distortion treatments when a user is presented with tasks in the space and the effectiveness of the distortion effect when using parts of the projected environment rather than the grid system. We hope to tackle this question in our future work. 

\section{Conclusion}
Through this project, we examined the Reality Distortion Room, a proof of concept that shows how visual perception of certain room distortion effects can invoke cohesive natural locomotion responses from the user. We tested a variety of spatial orientation visual effects. The user study demonstrated that the distortion treatments we designed were effective in influencing users’ natural locomotion in predictable ways. By relying on users' reactions to their visual perception of space, we can open new ways to engage with familiar environments, or to navigate in an enhanced, altered, or even completely virtual reality. We are especially excited that this study presents possibilities of instilling movement in people solely through visual deformations of the AR spaces they populate.

\acknowledgments{
The authors thank Emma Lin and Jacqueline Mei for their assistance, generosity, and continuous support in this project. The work was supported in part by NSF award IIS-2211784.
}

\bibliographystyle{abbrv-doi}

\bibliography{template}
\end{document}